\documentclass[aps,prl,twocolumn,superscriptaddress]{revtex4}
\usepackage{graphicx,color}
\usepackage{bm}
\usepackage{dsfont,amsmath,amssymb}
\usepackage{amsthm}

\newcommand{\bra}[1]{\mbox{$\langle #1 |$}}
\newcommand{\ket}[1]{\mbox{$| #1 \rangle$}}

\newcommand{\Tr}{\mathrm{Tr}}

\renewcommand{\Re}{\mathrm{Re}}
\renewcommand{\Im}{\mathrm{Im}}

\newcommand{\ee}{\mathrm{e}}
\newcommand{\ii}{\mathrm{i}}

\newcommand{\bl}{\color{black}}
\newcommand{\bk}{\color{black}}

\begin{document}
\title{Correlated and Critical Phenomena in Multipartite Quantum Non-Markovianity}
	\author{Ignacio González}
	\email{ignago10@ucm.es}
	\affiliation{Departamento de F\'{\i}sica Te\'orica, Facultad de Ciencias F\'isicas, Universidad Complutense, 28040 Madrid, Spain.}
	\author{\'Angel Rivas}
	\email{anrivas@ucm.es}
	\affiliation{Departamento de F\'{\i}sica Te\'orica, Facultad de Ciencias F\'isicas, Universidad Complutense, 28040 Madrid, Spain.}
	\affiliation{CCS-Center for Computational Simulation, Campus de Montegancedo UPM, 28660 Boadilla del Monte, Madrid, Spain.}

\date{\today}

\begin{abstract}
\bl We study non-Markovianity in the exact dynamics of two two-level atoms in a resonant cavity. We find a critical behavior in the form of a Markovian to non-Markovian transition at a finite interatomic distance and a discontinuity in the limit of infinitely close atoms. Actually, in this highly correlated regime, we find that non-Markovianity always arises if the system size exceeds a critical threshold. To this end, a non-singular, additive non-Markovian measure with appropriate asymptotic behavior is formulated. These results show how inter-subsystem distance and system size can act as precise controls for modulating memory effects in multipartite open quantum systems.\bk
\end{abstract}

\maketitle

\paragraph{Introduction.---} 
\bl Although Markovian master equations often approximate open quantum system dynamics effectively \cite{AlickiBook,BrPe02,Libro}, all open systems exhibit some degree of non-Markovianity, increasing in the presence of strongly coupled or structured environments. Understanding non-Markovian quantum systems is challenging, and its rigorous characterization has been subject to intense research, e.g. \cite{Lindblad79,Accardi82,BLP09,RHP10,Darek10,Darek11,Andrea14,Michael14,Darek-Sabrina, Kavan18, Streltsov20,Walter24}. However, most of the studies quantifying quantum non-Markovianity deal with relatively simple open systems, e.g. a qubit or an harmonic oscillator (see \cite{revMarko1,revMarko2,revMarko3} and references therein). The behavior of e.g. divisibility conditions or other non-Markovian characterizations has been addressed in very few multipartite cases \cite{Addis16,Benatti17,Loss24,Azevedo24,Benatti24,Sabale24,Keefe24,Meystre}. 

Interestingly, multipartite systems can exhibit non-Markovian dynamics even in unstructured environments  (e.g. flat spectral density). This has been studied, for example, in atomic arrays within 1D waveguides, where the environmental correlation function between separated atoms presents an effective width due to the finite speed of photons (retardation) \cite{Dinc, Meystre, Pichler, Sheremet}. In contrast, characterizing multipartite non-Markovianity in structured environments remains largely unexplored.

In this work, we advance in this direction by considering the non-Markovian dynamics of two atoms in a cavity separated by a distance $d$. By employing an improved non-Markovian quantifier and solving the dynamics in a numerically exact way, we find a novel, rich and singular, non-Markovian phenomenology, including:

i) A finite critical distance $d_c$ separating Markovian and non-Markovian regimes.

ii) An interval in $d$ where the non-Markovianity of the full dynamics is entirely due to the correlations between each individual atomic dynamics \cite{Markus1,Markus2}. 

iii) A finite distance $d_{\rm max}$ at which the amount of non-Markovianity is maximized in any bounded time interval $[0,t_{\rm end}]$, with a critical behavior in the form of a discontinuity in $d$ for $t_{\rm end}\to\infty$.

iv) Finally, a critical number of atoms above which the dynamics always becomes non-Markovian in the limit of very small $d$.

\bk

\paragraph{A non-Markovian measure for the multisystem case.---}
A number of different non-Markovianity measures have been proposed \cite{revMarko1,revMarko2,revMarko3}. One such measure \cite{RHP10} is given in terms of the trace norm of the Choi matrix of the intermediate dynamics:
\begin{equation}\label{RHP1}
\mathcal{N}_{\text{RHP}}(t):=\int_{0}^{t}g(s)ds, \quad g(t):=\lim_{\epsilon\downarrow0}\dfrac{\lVert \Upsilon_{(t+\epsilon,t)}\rVert_{1}-1}{\epsilon}.
\end{equation}
Here, $\Upsilon_{(t,s)}:=\left[\mathcal{E}_{(t,s)}\otimes\mathbb{I}\right]\ket{\Phi}\bra{\Phi}$, where $\mathcal{E}_{(t,s)}$ is given by the composition law $\mathcal{E}_{(t,0)}=\mathcal{E}_{(t,s)}\mathcal{E}_{(s,0)}$ with $t>s>0$, and $t=0$ is taken to be the initial time of the open system dynamics so that $\mathcal{E}_{(t,0)}$ is a completely positive \bl(CP) \bk and trace preserving map for all $t\geq0$. In addition, $\ket{\Phi}$ is the maximally entangled state between two copies of the system ${\ket{\Phi}=N^{-1/2}\sum_{n=1}^{N}\ket{n}\ket{n}}$ with $N$ the dimension of its Hilbert space, and $\lVert A\rVert_{1}=\text{Tr}\sqrt{A^{\dagger}A}$. \bl If $\mathcal{E}_{(t,s)}$ is CP for all $s\leq t \leq t_{\rm end}$, $\Upsilon_{(t,s)}$ is a density matrix and $\mathcal{N}_{\text{RHP}}(t_{\rm end})=0$. Thus, $\mathcal{N}_{\text{RHP}}$ assesses the deviation from CP-divisibility. \bk

For the multipartite case, and as a difference with other quantifiers, $\mathcal{N}_{\rm RHP}$ has the advantage of being stable under tensor products \cite{Benatti17,Benatti24}, i.e. if $\{\mathcal{E}_{(t,0)}^{(n)}\}_{n=1,2,\ldots}$ is a family of Markovian maps according to $\mathcal{N}_{\rm RHP}$, so is $\bigotimes_{n}\mathcal{E}_{(t,0)}^{(n)}$. In fact, a property which to our knowledge has not been highlighted before for $\mathcal{N}_{\rm RHP}$ is its additivity. To prove it, just note that $\log(x)\simeq x-1$ for $x\simeq 1$, so we can alternatively write $g(t)=\lim_{\epsilon\downarrow0}\log\lVert \Upsilon_{(t+\epsilon,t)}\rVert_1/\epsilon$, and, since the trace norm of the Choi matrix is multiplicative under tensor products (i.e. under independent dynamics), 
\begin{equation}\label{Choi_fact}
\mathcal{E}_{(t,0)}=\bigotimes_{n}\mathcal{E}^{(n)}_{(t,0)}\Rightarrow \lVert\Upsilon_{(\bl t,s\bk)}\rVert_{1}=\prod_{n}\lVert\Upsilon^{(n)}_{(\bl t,s\bk )}\rVert_{1}.
\end{equation}
The additivity and computational simplicity of $\mathcal{N}_{\text{RHP}}$ makes it an appealing choice to quantify non-Markovianity. However, its use also has some disadvantages. For instance, even if the dynamics is relaxing, i.e. the limit $\lim_{t\to\infty}\mathcal{E}_{(t,0)}=\mathcal{E}_\infty$ exists, $\mathcal{N}_{\text{RHP}}(t)$ might become arbitrarily large as $t$ increases. Such is the case in the known example of eternal non-Markovianity in \cite{Michael14}. For relaxing dynamics, it is natural to require that the non-Markovianity $\mathcal{N}$ approaches a limiting value, as no further memory effects can be observed once the system's evolution has essentially stopped near its relaxed state. \bl This effect can be accommodated in \eqref{RHP1} by introducing some weighted measure in the integral,  \bk

\begin{equation}\label{NMweight}
    \mathcal{N}=\int_0^{\infty}g(s)w_\mathcal{E}(s)ds,
\end{equation}
\bl where $w_{\mathcal{E}}(t)\geq0$ \bk and $\int_0^\infty w_{\mathcal{E}}(s)ds=1$, with $w_{\mathcal{E}}(t)$ approaching $w_\mathcal{E}(\infty)=0$ at a similar rate as $\mathcal{E}_{(t,0)}$ approaches $\mathcal{E}_\infty$. \bl However, for $\mathcal{E}_{(t,0)}=\mathcal{E}_{(t,0)}^{(1)}\otimes \mathcal{E}_{(t,0)}^{(2)}$, $\mathcal{N}$ is not additive because $g(t)$ already is, and $w_{\mathcal{E}}(t)$ decays within the relaxation time of the overall dynamics $\mathcal{E}_{(t,0)}$, which differs from the individual relaxation time of each $\mathcal{E}_{(t,0)}^{(j)}$ unless they are identical. Thus, we can at most require additivity of $\mathcal{N}$ for copies of the same dynamics by considering the natural condition $w_{(\mathcal{E}^{\otimes n})}(t)=w_{\mathcal{E}}(t)$. 

\bk A sensible choice which fulfills these requirements is
\begin{equation}\label{logweight}
    w_{\mathcal{E}}(t)=\dfrac{\log F(t)}{\int_{0}^{\infty}\log F(s)ds}.
\end{equation}
Here, $F(t):=\|\sqrt{\Upsilon_{(t,0)}}\sqrt{\Upsilon_{\infty}}\|_1^2$ is the fidelity between the Choi matrix at time $t$ and asymptotically $\Upsilon_{\infty}:=\lim_{t\to\infty}\Upsilon_{(t,0)}$ (for non-relaxing dynamics one can take $\Upsilon_{\infty}:=\lim_{T\to\infty}\frac{1}{T}\int_0^{T}\Upsilon_{(t,0)}dt$). Note that the introduction of the weight function gives $\mathcal{N}$ dimensions of $[\rm time]^{-1}$. Since $-\int_{0}^{\infty}\log F(s)ds$ is a logarithmic estimate of the relaxation time of the dynamics, $\mathcal{N}$ in \eqref{NMweight} becomes a quantifier of non-Markovianity per unit of time during which there is significant evolution.

Another issue with the quantifier \eqref{RHP1} has to do with the fact that $g(t)$ can be unbounded and non-integrable even in finite intervals. Such is the case of the damped Jaynes-Cummings model \bl under a Lorentzian spectral density \bk
\begin{equation}\label{spectraldensity}
    J(\omega)=\frac{1}{2\pi}\frac{\gamma_0^2 \lambda}{(\omega-\omega_0)^2+\lambda^2},
\end{equation}\bl
which is used to describe the interaction of a two-level atom with a lossy resonant cavity \cite{Garraway97,GKnight,Kurizki}. Here, $\omega_0$ is the resonant frequency (we shall take units of $\hbar=c=1$ from now on), and $\gamma_0$ and $\lambda$ account for the coupling strength and the frequency width of the environment\bk, respectively. This model can be solved exactly for $\omega_0/\lambda$ large enough to justify the frequency range extension to negative values without introducing significant errors \cite{BrPe02,Garraway97}, obtaining the following exact master equation
\begin{equation}
    \frac{d\rho(t)}{dt}=-\ii[H_S,\rho(t)]+\gamma(t)\big[\sigma_-\rho\sigma_+-\tfrac12\{\sigma_+\sigma_-,\rho(t)\}\big].
\end{equation}
Here, $H_S=\omega_0\sigma_z/2$ is the free Hamiltonian of the atom with ground and excited states, $|0\rangle$ and $|1\rangle$, respectively, $\sigma_z=[\sigma_+,\sigma_-]$, $\sigma_+=|1\rangle\langle 0|=\sigma_-^\dagger$, and the single decay rate is
\begin{equation}\label{dr1atom}
    \gamma(t)=\dfrac{2\gamma_0^2}{\lambda+\Omega_{1}\coth\left(\frac{\Omega_{1} t}{2}\right)},
\end{equation}
with $\Omega_{1}=\sqrt{\lambda^2-2\gamma_0^2}$. The function $g(t)$ can be easily computed resulting in the negative part of $\gamma(t)$,  \begin{equation}\label{g1atom}
g_{\rm atom}(t)=\gamma^-(t):=\tfrac{1}{2}\big[\lvert\gamma(t)\rvert-\gamma(t)\big].  
\end{equation}
It is clear that the dynamics is non-Markovian only if ${\lambda<\sqrt{2}\gamma_0}$, in which case $g(t)$ is a periodic function with simple poles at $t_n=\frac{2}{|\Omega_1|} [\pi  n-\cot^{-1}(\tfrac{\lambda}{|\Omega_1|})]$, with $n=1,2,\ldots$ In order to regularize the integral \eqref{NMweight} while preserving the additive property, we can apply a ``square root-power'' procedure
\begin{equation}\label{NMweightsqrt}
    \tilde{\mathcal{N}}=\left[\int_0^{\infty}\sqrt{g(s)}w_{\mathcal{E}}(s)ds\right]^2.
\end{equation}
Accordingly, for some $g(t)$ with a pole of order $\alpha$, we can take the ($\alpha+1$)th root followed by an ($\alpha+1$)th power to regularize \eqref{NMweight} keeping additivity under copies of the same dynamics. Note that other regularizations valid for any kind of pole \cite{revMarko1} e.g. $\bar{g}(t)\propto\tanh[\eta_0 g(t)]$ (with $\eta_0>0$ a constant) come at the price of the additivity of the measure.

In all these integrals the upper limit can be substituted by a finite final time $t_{\rm end}$ physically motivated by the experimental setup.

\begin{figure}[t]
	\includegraphics[width=\columnwidth]{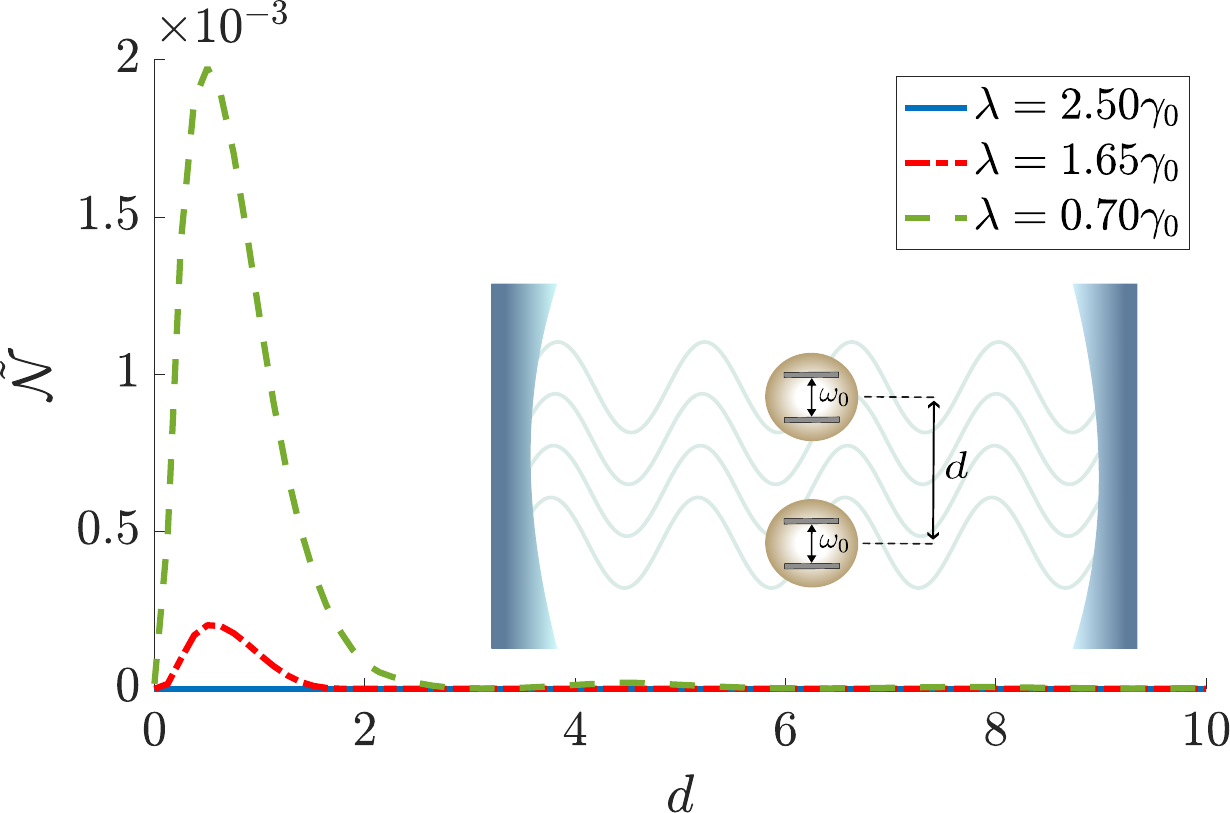}
	\caption{Non-Markovianity measure \eqref{NMweightsqrt} in the time interval $[0,350]$ as a function of $d$ for different values of $\lambda$ ($\gamma_0=0.01$). All quantities are in units of $\omega_0=1$. The insert illustrates the system setup.}
	\label{Fig:1}
\end{figure}

\paragraph{Two atoms in a resonant QED cavity.---}
Consider now a pair of two-level atoms with free Hamiltonian $H_S=\frac{\omega_0}{2}(\sigma_z^{(1)}+\sigma_z^{(2)})$ inside a resonant QED cavity (see Fig. \ref{Fig:1}) modelled by the Hamiltonian $H_B=\sum_{\bm{k}} \omega_k a^\dagger_{\bm{k}}a_{\bm{k}}$, with $\omega_k=|\bm{k}|$ and $a_{\bm{k}}$ the annihilation operator for a photon with wavevector $\bm{k}$. The interaction Hamiltonian under the rotating wave approximation \bl (RWA) \bk in the interaction picture is given by \bl \cite{Garraway97,GKnight,Milonni,Wolf} \bk 
\begin{equation}\label{hamiltonian2at}
		H_{I}(t)=\displaystyle\sum_{\bm{k}}g(\omega_{k})\left[\sum_{n=1,2}\sigma_{+}^{(n)}a_{\bm{k}}\ee^{\ii\bm{k}\cdot\bm{r}_{n}} \ee^{-\ii\Delta_{k}t}+\text{H.c.}\right],
\end{equation}
where $\bm{r}_{n}$ is the position of the $n$th atom, $\sigma_{+}^{(n)}={\ket{1}_{n}\!\bra{0}}$, and $\Delta_{k}=\omega_{k}-\omega_0$. If we assume that the cavity remains initially in the vacuum \bl and at most one excitation is injected into the system, since $H_{I}(t)$ preserves the total number of excitations, we can restrict the dynamics to the (0 and) 1-excitation subspace. Then \bk the system+cavity state at a time $t$ is of the form 
\begin{align}\label{state}
\ket{\psi(t)}=\sum_{i+j<2}c_{ij}(t)\ket{i,j,0}+\sum_{\bm{k}}c_{00\bm{k}}(t)\ket{0,0,1_{\bm{k}}},
\end{align}
where $|i,j,m_{\bm{k}}\rangle$ indicates the first and second atom in the states $\ket{i}$ and $\ket{j}$, respectively, and $m$ photons with wavevector $\bm{k}$ in the bath, $i,j,m \in\{0,1\}$. The first component is clearly invariant, $\dot{c}_{00}(t)=0$, so we shall not consider it in the following. The Schrödinger equation on the state \eqref{state} gives
\begin{equation}\label{schro}
		\begin{cases}
			\displaystyle \ii\dot{c}_{jl}(t)=\sum_{\bm{k}}g(\omega_{k})\ee^{\ii\bm{k}\cdot\bm{r}_{1+l}}\ee^{-\ii\Delta_{k}t}c_{00\bm{k}}(t),\\
			  \ii\dot{c}_{00\bm{k}}(t)=g(\omega_{k})\ee^{\ii\Delta_{k}t} \left[\ee^{-\ii\bm{k}\cdot\bm{r}_1}c_{10}(t)+\ee^{-\ii\bm{k}\cdot\bm{r}_2}c_{01}(t)\right].		
		\end{cases}
\end{equation}
Since $c_{00\bm{k}}(0)=0$, integrating formally the last equation and introducing the result into the first one yields
\begin{equation}\label{intdiff1}
    \dot{c}_{jl}(t)=-\int_0^t dt' \left[f_{1}(t-t')c_{jl}(t')+f_{2}(t-t')c_{lj}(t')\right],
\end{equation}
where we have defined the correlation functions, after taking the continuum limit $\sum_{\bm{k}}\rightarrow \lim_{V\to\infty}V/(4\pi^3
)\int\,d^3k$ (with $V$ the quantization volume), by
\begin{align}
        f_{1}(t-t'):&=\int_{0}^{\infty}d\omega\,J(\omega)\ee^{-\ii\Delta(t-t')}, \label{corr_def1}\\
        f_{2}(t-t'):&=\int_{0}^{\infty}d\omega\,J(\omega)\dfrac{\sin\left(\omega d\right)}{\omega d}\ee^{-\ii\Delta(t-t')}\label{corr_def2},
\end{align}
with $\Delta=\omega-\omega_0$, $d=\lvert\bm{r}_{1}-\bm{r}_2\rvert$ and $J(\omega):=\frac{\omega^2}{\pi^2}\lim_{V\to\infty}g^2(\omega)V$ the spectral density. As usual for modeling a resonant cavity, we consider the Lorentzian spectral density as in the previous section \eqref{spectraldensity},  \bl with $\gamma_0\ll \omega_0 $ (required for the RWA that fails for strong coupling) \bk and $\omega_0/\lambda$ large enough to extend the integrals in \eqref{corr_def1}-\eqref{corr_def2} to negative frequencies. In this case one obtains
\begin{equation}\label{corr11}
	f_{1}(t)=\dfrac{\gamma_0^2}{2}\ee^{-\lambda t},
\end{equation}	
\begin{align}\label{corr12}
f_{2}(t)=\dfrac{\gamma_0^2}{4d}&\left\{\left[\frac{\theta(t-d)\ee^{(\lambda+\ii\omega_0)d}-\ee^{-(\lambda+\ii\omega_0)d}}{\lambda+\ii\omega_0}\right]\ee^{-\lambda t}\right.\nonumber\\
&-\left.\theta(d-t)\left[\frac{\ee^{-(\lambda-\ii\omega_0)d}}{\lambda-\ii\omega_0}\ee^{\lambda t}-\frac{2\lambda}{\omega_0^2+\lambda^2}\ee^{\ii\omega_0t}\right]\right\}.
\end{align}

\paragraph{Exact master equation for the reduced system.---} 
If we write the  atomic part of the state \eqref{state} in terms of $\ket{\pm}=\frac{1}{\sqrt{2}}(|10\rangle\pm|01\rangle)$, and trace out the bath's degrees of freedom, after time differentiation we obtain that the atomic dynamics satisfies the master equation (see Appendix A)
\begin{align}\label{masterequation}
\dfrac{d\rho_{S}(t)}{dt}&=-\ii\sum_{m=1,2}\dfrac{S_{m}(t)}{2}\left[L_{m}^{\dagger}L_{m},\rho_{S}(t)\right]\nonumber\\
&+\sum_{m=1,2}\gamma_{m}(t)\left[L_{m}\rho_{S}(t)L_{m}^{\dagger}-\dfrac{1}{2}\left\{L_{m}^{\dagger}L_{m},\rho_{S}(t)\right\}\right],
\end{align}
with the jump operators $L_{m}=\frac{1}{\sqrt{2}}[\sigma_{-}^{(1)}-(-1)^{m}\sigma_{-}^{(2)}]$, and Lamb shifts and decay rates given by
\begin{equation}
S_{m}(t):=-2\,\Im \bigg[\frac{\dot{r}_{m}(t)}{r_{m}(t)}\bigg],\quad \gamma_m(t)=-2\, \Re \bigg[\frac{\dot{r}_m(t)}{r_m(t)}\bigg],
\end{equation}
where $r_m(t):=\frac{1}{\sqrt{2}}[c_{10}(t)-(-1)^{m}c_{01}(t)]$. These, according to \eqref{intdiff1}, satisfy
\begin{equation}\label{intdiff2}
    \dot{r}_{m}(t)=-\int_{0}^{t}dt'K_{m}(t-t')r_{m}(t'), 
\end{equation}
with $K_{m}(t):=f_1(t)-(-1)^mf_2(t)$.

\paragraph{Non-Markovianity as a function of interatomic distance.---} 
The function $g(t)$ can be calculated from the master equation \eqref{masterequation} \cite{revMarko1,Michael14}, yielding
\begin{equation}\label{g2atoms}
g(t)=\frac{2}{3} \big[\gamma_1^-(t)+\gamma_2^-(t)\big].
\end{equation}
To obtain the decay rates we need to solve \eqref{intdiff2}. This is difficult because of the intricate time dependence of $f_2(t)$ for a finite distance $d$, Eq.~\eqref{corr12}. However, when $d\rightarrow\infty$, we have $f_{2}(t)=0$, hence $K_{1}(t)=K_{2}(t)=f_{1}(t)$. The system \eqref{intdiff2} reduces to that found in the one-atom damped Jaynes-Cummings model, and we obtain $\gamma_1(t)=\gamma_2(t)=\gamma(t)$,
\begin{equation}\label{g2atomsInf}
		g_{\infty}(t)=\frac{4}{3} \gamma^-(t)=\dfrac{4}{3}g_{\text{atom}}(t),
\end{equation}
where $\gamma(t)$ and $g_{\text{atom}}(t)$ are given by \eqref{dr1atom} and \eqref{g1atom}, respectively. One should notice that, despite having independent atomic dynamics in the limit $d\to \infty$, as we have restricted our study to the (0 and) 1 excitation subspace, the additivity factor of $2$ is here reduced to $4/3$. On the other hand, if we take $d=0$, $f_{1}(t)=f_{2}(t)$, so that $K_{1}(t)=2f_1(t)$ and $K_{2}(t)=0$. The equations \eqref{intdiff2} can be solved by differentiation leading to \cite{2atd0,2atd02}
\begin{equation}
        r_{1}(t)=\ee^{-\frac{\lambda t}{2}}\left[\cosh\left(\dfrac{\Omega_{2}t}{2}\right)+\dfrac{\lambda}{\Omega_{2}}\sinh\left(\dfrac{\Omega_{2}t}{2}\right)\right],\\
\end{equation}
and $r_{2}(t)=r_{2}(0)$, with $\Omega_{2}=\sqrt{\lambda^2-4\gamma_0^2}$. Therefore $\ket{-}$ becomes a dark state and jointly with $\ket{0,0}$ forms a decoherence-free subspace (DFS) \cite{DFS1,DFS2}. In this case $\gamma_2(t)=0$ and $\gamma_1(t)$ is twice the same function as \eqref{dr1atom} with $\Omega_2$ in the role of $\Omega_1$, so that
\begin{equation}\label{gd0}
		g_{0}(t)=\frac{2}{3} \gamma_1^-(t)=\frac{8\gamma_0^2}{3}\left[\dfrac{1}{\lambda+\Omega_{2}\coth\left(\frac{\Omega_{2} t}{2}\right)}\right]^-.
\end{equation}
Therefore, when ${d=0}$ the dynamics is non-Markovian for $\lambda<2\gamma_0$. In the opposite limit $d\to\infty$, with $g_\infty(t)$ being proportional to that of the one-atom case \eqref{g2atomsInf}, the dynamics is non-Markovian when $\lambda<\sqrt{2}\gamma_0$. Consequently, for $2\gamma_0>\lambda>\sqrt{2}\gamma_0$ the dynamics is Markovian when the atoms are far from each other, but non-Markovian when they are at the same position, which indicates that non-Markovianity can be created by bringing the atoms together. The question is how close the atoms must be to observe this effect.

\begin{figure}[t]
	\includegraphics[width=\columnwidth]{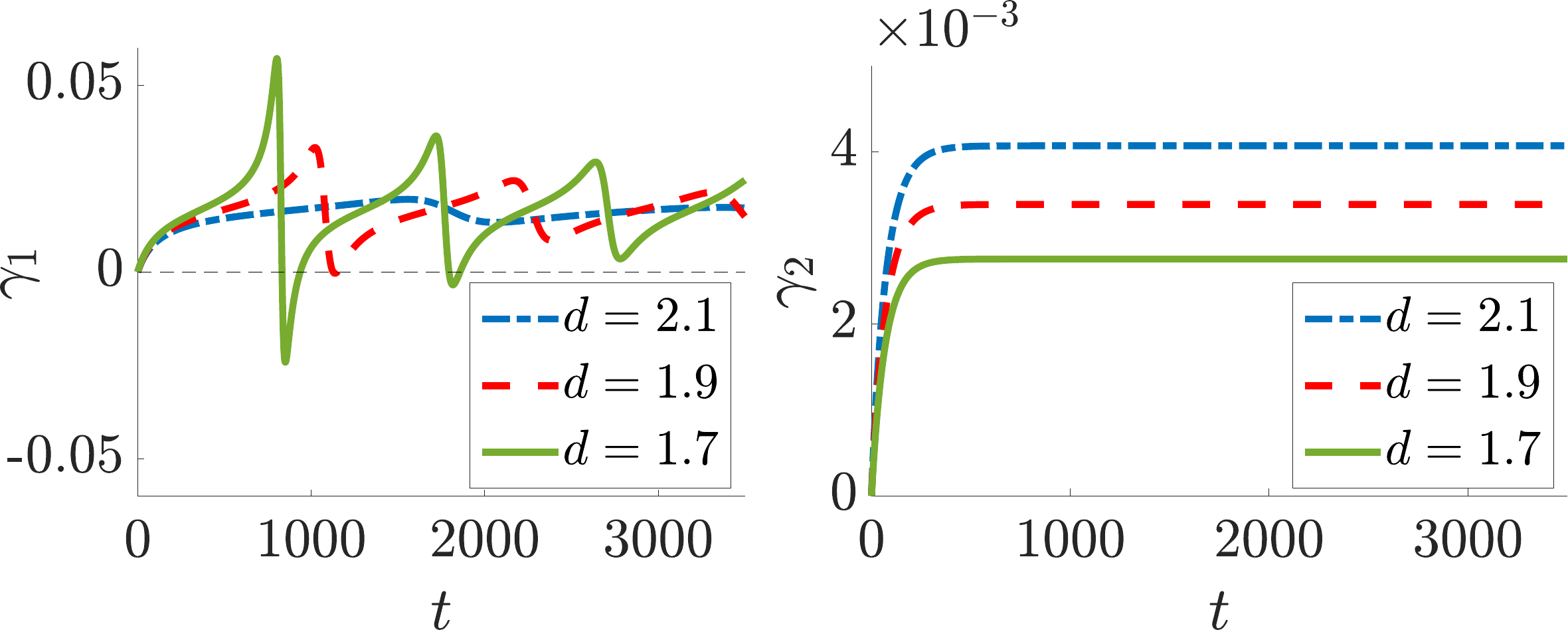}
	\caption{Decay rates $\gamma_1(t)$ (left) and $\gamma_2(t)$ (right) for different values of $d$ (units of $\omega_0=1$). For $d=2.1$ both decay rates are always positive. As $d$ is decreased, $\gamma_{1}(t)$ begins to dip below the $x$ axis. We have taken $\gamma_0=0.01$, and $\lambda=1.65\gamma_0$.}
	\label{Fig:3}
\end{figure}
To answer such a question we need to solve the dynamics for any distance $d$. The intricate time dependence of $f_2(t)$ prevents an analytical solution as well as the application of available non-Markovian numerical methods \cite{HEOM1,HEOM2,Tamascelli,Lambdert19,GPseudomodes,TEMPO} that are efficient on bath correlation functions well approximated by linear combinations of exponential terms and/or with short correlation time (i.e. large $\lambda$). Nevertheless, \eqref{intdiff2} defines a pair of uncoupled integro-differential equations of Volterra type with a continuous kernel, which can be numerically solved by quadrature rules \cite{Book_Linz85,Book_Delves85}. The numerical results for the decay rates $\gamma_{1,2}(t)$ are plotted in Fig.~\ref{Fig:3} for different distances confirming the existence of a finite critical distance $d_{c}>0$ where the minimum of $\gamma_1(t)$ becomes negative. As illustrated in Fig.~\ref{Fig:1}, the dynamics remains Markovian at any distance in the region $\lambda>2\gamma_0$. \bl For $2\gamma_0>\lambda>\sqrt{2}\gamma_0$, we see non-Markovianity being created at $d_c$ as commented. Finally, if $\lambda<\sqrt{2}\gamma_0$  we find that the dynamics is non-Markovian at any distance, $\tilde{\mathcal{N}}\rvert_{d}>0$. 

\bl It is interesting to consider the uncorrelated part of the dynamics by removing the crossed terms between different atoms in \eqref{masterequation}, calculate $\tilde{\mathcal{N}}$ in that case, and compare with the total value (see Appendix B). 
This is done in Fig.~\ref{Fig:2}. For $2\gamma_0>\lambda>\sqrt{2}\gamma_0$, the uncorrelated value for the non-Markovianity is 3 orders of magnitude smaller than the total one, vanishing for $d$ larger than some critical value $d_{\rm uc}$, with $d_{\rm uc}<d_{c}$. Therefore, if $d\in(d_{\rm uc},d_{c})$ the non-Markovianity of the system is entirely due to the correlations among its parts. For $\lambda<\sqrt{2}\gamma_0$, the non-Markovianity is also significantly reduced for the uncorrelated dynamics, unless for very small $d$. For $\lambda>2\gamma_0$, both the uncorrelated part and the total dynamics remain Markovian at any distance. \bk

\begin{figure}[t]
	\includegraphics[width=\columnwidth]{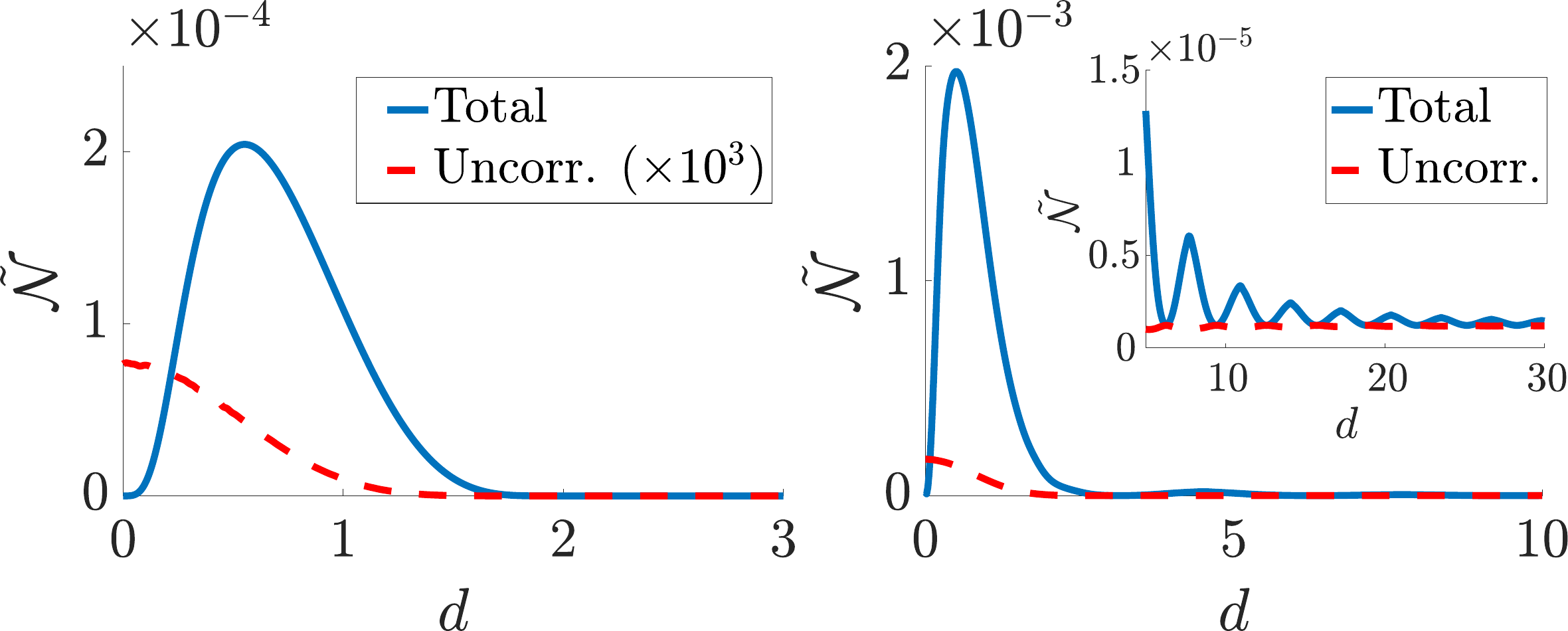}
	\caption{\bl Comparison between the non-Markovianity measure for the uncorrelated and total dynamics in the time interval $[0,350]$ as a function of $d$. For $\lambda=1.65\gamma_0$ (left), $d_{\rm c}\simeq 1.904$ and $d_{\rm uc}\simeq 1.870$. For $\lambda=0.7\gamma_0$  (right), both curves approach the value $\tilde{\mathcal{N}}|_{d\rightarrow\infty}=1.2112\cdot 10^{-6}$ at large $d$. All quantities are in units of $\omega_0=1$, and $\gamma_0=0.01$.\bk}
	\label{Fig:2}
\end{figure}

\paragraph{Maximum non-Markovianity.---}
Interestingly, the amount of non-Markovianity as measured by \eqref{NMweightsqrt} presents a non-monotonic behavior with a maximum at an intermediate $d_{\rm max}$ distance as show in Fig.~\ref{Fig:1}. This profile can be understood as a result of the slowing down of the dynamics due to the DFS existing at  $d=0$. Specifically, as the atoms get closer, the relaxation time $\tau_2$ of the singlet component $r_{2}(t)$ increases, approaching infinity as $d\rightarrow0$. Consequently, we can consider three different regimes for the non-Markovianity $\tilde{\mathcal{N}}(t_{\rm end})$, with $t_{\rm end}$ any finite value for the upper limit in the integral \eqref{NMweightsqrt}. For large $d$, $\tau_2\ll t_{\rm end}$, and the weight function is negligible in $(\tau_2,t_{\rm end}]$ and these times do not contribute to the integral \eqref{NMweightsqrt}. As $d$ is decreasing, the system eventually reaches the regime where $\tau_2\sim t_{\rm end}$ and the non-Markovianity increases because the weight function is non-negligible over the total integration interval $[0,t_{\rm end}]$, and the $g(t)$ function grows as the negative part of $\gamma_1(t)$ increases. Finally, for very small $d$, $\tau_2\gg t_{\rm end}$, the singlet component decays so slowly that it remains approximately constant $r_{2}(t)\simeq r_2(0)\simeq r_{2}(t_{\rm end})$, and so the system effectively relaxes in the same time scale $\tau_1$ as $r_1(t)$. This, in turn, results in the weight function falling to 0 at an earlier time $\tau_1$($<\tau_2$), leading to a decreasing value of $\tilde{\mathcal{N}}(t_{\rm end})$ that continuously approaches $\tilde{\mathcal{N}}(t_{\rm end})\rvert_{d=0}$. This illustrates the convenience of using the weight function in the definition of $\tilde{\mathcal{N}}$: for the same finite time interval, the non-Markovianity decreases in the limit of extremely slow dynamics.

\begin{figure}[t]
	\includegraphics[width=\columnwidth]{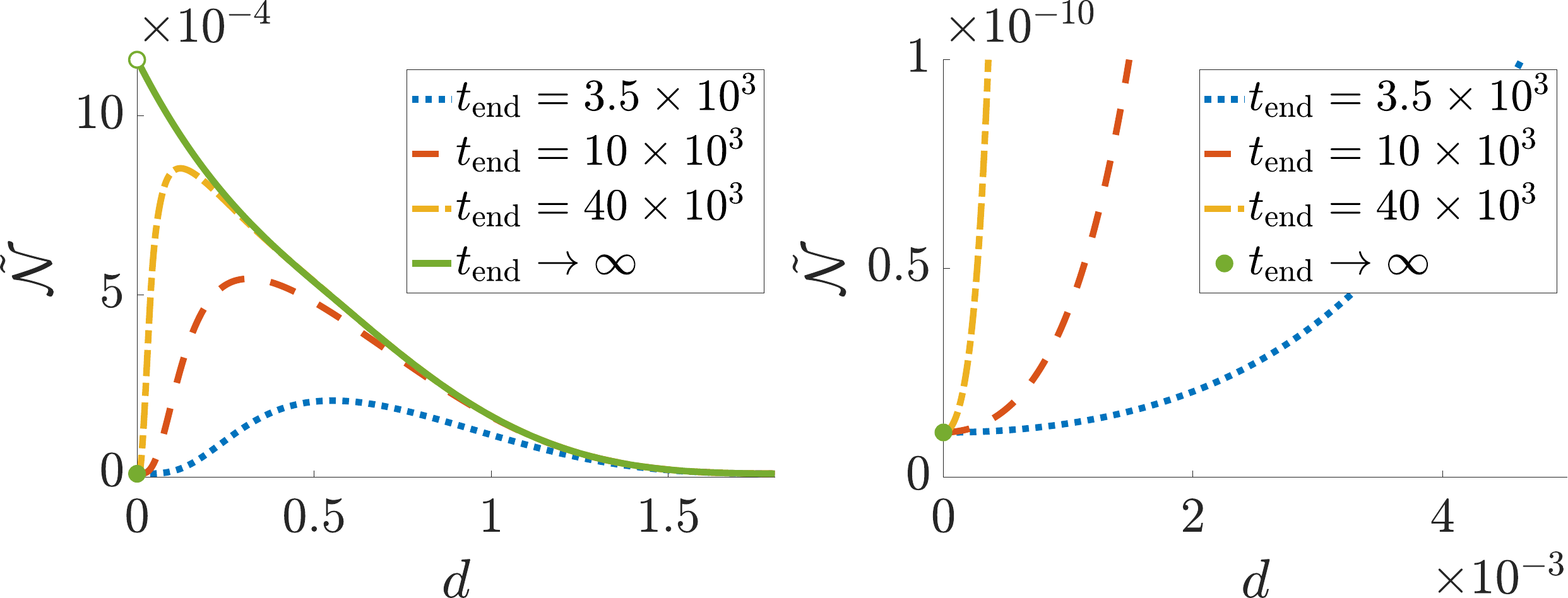}
	\caption{Non-Markovianity measure \eqref{NMweightsqrt} as a function of $d$ for different values of $t_{\text{end}}$ (units of $\omega_0=1$). The plot on the right shows the neighborhood of $d=0$. We have taken the same values of $\gamma_0$ and $\lambda$ as in Fig. \ref{Fig:3}. The line $t_{\rm end}\to\infty$ is an extrapolation from the exact value at $d\downarrow0$ and $t_{\rm end}=1.2\times 10^4$. Note the discontinuity of the green line for $d=0$.}
	\label{Fig:4}
\end{figure}

The values of $d$ which lead to an approximately constant singlet component are smaller as $t_{\rm end}$ increases. Therefore, the peak moves to the left and becomes higher as $t_{\rm end}$ grows due to the broader integration range. In the limit $t_{\rm end}\rightarrow\infty$, $g(t)$ approaches the periodic function $g_0(t)$ and it can be shown (see Appendix C) 
that $\lim_{d\downarrow0}\tilde{\mathcal{N}}\rvert_{d}(t_{\rm end}\to\infty)\equiv\lim_{d\downarrow0}\tilde{\mathcal{N}}\rvert_{d}=\frac{1}{3}(2\gamma_0-\lambda)$. However, at $d=0$ $r_{2}(t)$ remains constant for all times, and the system relaxes after a time $\tau_1$, when $r_{1}(t)$ does. So, the weight function changes discontinuously and decays in a time $\tau_1$, as a difference with the limiting case $d\downarrow0$ where it takes an infinite time to decay due to the contribution of $\tau_2$. Therefore the non-Markovianity is suddenly modified at $d=0$ compared to the limiting case $d\downarrow0$, $\tilde{\mathcal{N}}\rvert_{d=0}\neq\lim_{d\downarrow0}\tilde{\mathcal{N}}\rvert_{d}$ (see Fig.~\ref{Fig:4}). This discontinuity at $d=0$ is a critical behaviour in the non-Markovianity caused by the breaking of the symmetry $\bm{r}_1\leftrightarrow\bm{r}_2$ in $H_I(t)$ that leads to the formation of the DFS. \bl Note that by considering only the uncorrelated part of the dynamics, there is no such symmetry, no DFS is formed, and the maximum non-Markovianity is obtained at $d=0$ instead of $d_{\rm max}$, as shown in Fig.~\ref{Fig:2}.\bk

\paragraph{Non-Markovianity of $N$ atoms at $d=0$.---} 
Non-Markovianity can be influenced also by modifying the number of atoms in the cavity. More specifically, consider the previous situation with $N$ atoms at $d=0$. The interaction Hamiltonian \eqref{hamiltonian2at} now becomes
\begin{equation}\label{hamiltonianNat}
    H_{I}(t)=\displaystyle\sum_{\bm{k}}g(\omega_{k})\sum_{n=1}^{N}\left[\sigma_{+}^{(n)}a_{\bm{k}}\ee^{-\ii\Delta_{k}t}+\text{H.c.}\right].
\end{equation}
By restricting again to the (0 and) 1 excitation subspace, and generalizing the arguments used for $N=2$, we obtain (see Appendices D and E) 
the following master equation for the $N$ atoms
\begin{equation}\label{meNat}
	\dfrac{d\rho_{S}(t)}{dt}=\gamma_{N}(t)\left(J_{-}\rho_{S}(t)J_{+}-\dfrac{1}{2}\left\{J_{+}J_{-},\rho_{S}(t)\right\}\right),
\end{equation}
with the jump operator $J_{-}=J_{+}^{\dagger}=\frac{1}{\sqrt{N}}\sum_{n=1}^{N}\sigma_{-}^{(n)}$, and the decay rate
\begin{equation}
	\gamma_{N}(t)=\dfrac{2N\gamma_0^2}{\lambda+\Omega_{N}\coth\left(\frac{\Omega_{N}t}{2}\right)},
\end{equation}
where $\Omega_{N}=\sqrt{\lambda^2-2N\gamma_0^2}$. In this case, we obtain
\begin{equation}
	g_{(N)}(t)=\dfrac{2}{N+1}\gamma_{N}^-(t).
\end{equation}
Therefore, the dynamics is non-Markovian for $\lambda<\sqrt{2N}\gamma_0$. Thus, a Markovian to non-Markovian transition can be induced by increasing the number of atoms beyond $N_{c}:=\lceil\frac{\lambda^2}{2\gamma_0^2}\rceil$. Particularly, in the thermodynamic limit $N\to\infty$ the dynamics becomes non-Markovian for any finite $\lambda$ and $\gamma_0$. 

\paragraph{Conclusions.---} 
\bl We have characterized the dynamics of two two-level atoms inside a resonant cavity as a function of their spatial separation with the help of an improved non-Markovianity measure, yielding several significant findings. We have identified, for the first time to our knowledge, a finite critical interatomic distance $d_c$ separating Markovian from non-Markovian dynamics, with Markovian behavior at large distances. We also found that the non-Markovianity can be entirely due to correlations among the dynamics of each atom. In addition, a finite distance $d_{\rm max}$ at which non-Markovianity reaches a maximum has been identified, with a discontinuity at $d=0$ for $t_{\rm end}\to\infty$ due to a symmetry breaking in the non-Markovian dynamics. Finally, we also obtain a critical number of atoms above which the dynamics always becomes non-Markovian in the limit of vanishing separation (where all retardation effects vanish), suggesting that Markovian approximations should be cautiously applied when considering large systems. Importantly, these results are independent of our particular choice of weight function in \eqref{NMweightsqrt}.

Note that this phenomenology differs substantially from existing studies on atoms in 1D waveguides because of the 3D structured environment considered here, which allows, for example, the emergence of Markovian dynamics at large distances. Furthermore, it illustrates how parameters, such as inter-subsystem distance and system size, can act as precise controls for modulating memory effects. This tunability provides a practical route to engineer non-Markovian dynamics in quantum platforms like atomic and ionic traps, eliminating the need for intricate environmental modifications typically unfeasible in experimental settings.

\bk

\begin{acknowledgments}
The authors acknowledge support from Spanish MICIN grant PID2021-122547NB-I00 and from the “MADQuantum-CM" project funded by Comunidad de Madrid and by the Recovery, Transformation and Resilience Plan – Funded by the European Union - NextGenerationEU. I.G. acknowledges support from the MICIN contract PRE2022-101824 (MICIN/AEI/FSE,UE).
\end{acknowledgments}

\setcounter{secnumdepth}{2}
\setcounter{equation}{0}
\setcounter{figure}{0}
\renewcommand{\theequation}{A\arabic{equation}}
\renewcommand{\thetable}{A\arabic{table}}

\appendix

\section{Exact master equation for two atoms at different positions} 
The total system+environment state at an arbitrary time $t$, 
\begin{multline}
    \ket{\psi(t)}=c_{00}\ket{0,0,0}+c_{10}(t)\ket{1,0,0}\\
    +c_{01}(t)\ket{0,1,0}+\sum_{\bm{k}}c_{00\bm{k}}(t)\ket{0,0,1_{\bm{k}}},
\end{multline}
is rewritten in the basis $\{\ket{00},\ket{\pm}=\frac{1}{\sqrt{2}}\left(\ket{10}\pm\ket{01}\right)\}$ as
\begin{multline}\label{state_std}
\ket{\psi(t)}=c_{00}\ket{0,0,0}+r_{1}(t)\ket{+}\ket{0}\\
+r_{2}(t)\ket{-}\ket{0}+\sum_{\bm{k}}c_{00\bm{k}}(t)\ket{0,0,1_{\bm{k}}},
\end{multline}
with
\begin{equation}\label{comp_std_coeff}
	\begin{cases}
		r_{1}(t)=\dfrac{1}{\sqrt{2}}\left[c_{10}(t)+c_{01}(t)\right],\\
	 	r_{2}(t)=\dfrac{1}{\sqrt{2}}\left[c_{10}(t)-c_{01}(t)\right]. 
	\end{cases}
\end{equation}
The reduced system density matrix can then be easily calculated, resulting in
\begin{multline}\label{rho}
\rho_{S}(t)=\text{Tr}_{B}\big[\ket{\psi(t)}\bra{\psi(t)}\big]\\
=\begin{pmatrix}
		\lvert r_{1}(t)\rvert^2 & r_{1}(t)r_{2}^{*}(t) & r_{1}(t)c_{00}^{*}\\[10pt]
		r_{1}^{*}(t)r_{2}(t) & \lvert r_{2}(t)\rvert^2 & r_{2}(t)c_{00}^{*}\\[10pt]
		r_{1}^{*}(t)c_{00} & r_{2}^{*}(t)c_{00} & 1 - \lvert r_{1}(t)\rvert^2-\lvert r_{2}(t)\rvert^2 
	\end{pmatrix}.
\end{multline}
After taking time-derivative, we obtain
\begin{widetext}
\begin{align}\label{masterequationLH}
\dfrac{d\rho_{S}(t)}{dt}=&-\ii\,\text{Im}\left[\dfrac{\dot{r}_{1}(t)}{r_{1}(t)}\right]\begin{pmatrix}
		0 & -r_{1}(t)r_{2}^{*}(t) & -r_{1}(t)c_{00}^{*}\\[10pt]
		r_{1}^{*}(t)r_{2}(t) & 0 & 0\\[10pt]
		r_{1}^{*}(t)c_{00} & 0 & 0
	\end{pmatrix}
	-\ii\,\text{Im}\left[\dfrac{\dot{r}_{2}(t)}{r_{2}(t)}\right]\begin{pmatrix}
		0 & r_{1}(t)r_{2}^{*}(t) & 0\\[10pt]
		-r_{1}^{*}(t)r_{2}(t) & 0 & -r_{2}(t)c_{00}^{*}\\[10pt]
		0 & r_{2}^{*}(t)c_{00} & 0
	\end{pmatrix}\nonumber \\[10pt]
 	&+\text{Re}\left[\dfrac{\dot{r}_{1}(t)}{r_{1}(t)}\right]\begin{pmatrix}
		2\lvert r_{1}(t)\rvert^2 & r_{1}(t)r_{2}^{*}(t) & r_{1}(t)c_{00}^{*}\\[10pt]
		r_{1}^{*}(t)r_{2}(t) & 0 & 0\\[10pt]
		r_{1}^{*}(t)c_{00} & 0 & -2\lvert r_{1}(t)\rvert^2
	\end{pmatrix}
 +\text{Re}\left[\dfrac{\dot{r}_{2}(t)}{r_{2}(t)}\right]\begin{pmatrix}
		0 & r_{1}(t)r_{2}^{*}(t) & 0\\[10pt]
		r_{1}^{*}(t)r_{2}(t) & 2\lvert r_{2}(t)\rvert^2 & r_{2}(t)c_{00}^{*}\\[10pt]
		0 & r_{2}^{*}(t)c_{00} & -2\lvert r_{2}(t)\rvert^2
	\end{pmatrix}.
\end{align}
Note that the fractions $\dot{r}_{m}(t)/r_{m}(t)$ are independent of the initial system state. Indeed, as $r_m(t)$ satisfy a homogeneous integro-differential equation (see main text)
\begin{equation}
    \dot{r}_{m}(t)=-\int_{0}^{t}dt'K_{m}(t-t')r_{m}(t'), 
\end{equation}
if $\tilde{r}_m(t)$ is the solution for the initial condition $\tilde{r}_m(t)=1$, the solution for an arbitrary initial condition $r_m(0)$ is clearly $r_m(t)=\tilde{r}_m(t)r_m(0)$, so the $\dot{r}_{m}(t)/r_{m}(t)$ are independent of $r_m(0)$.

On the other hand, if we define the jump operators
\begin{equation}\label{jump}
	\begin{cases}
		L_{1}=\ket{0,0}\bra{+}=\dfrac{1}{\sqrt{2}}\left(\sigma_{-}^{(1)}+\sigma_{-}^{(2)}\right),\\[8pt]
		L_{2}=\ket{0,0}\bra{-}=\dfrac{1}{\sqrt{2}}\left(\sigma_{-}^{(1)}-\sigma_{-}^{(2)}\right),
	\end{cases}
\end{equation}
it is straightforwardly shown that
\begin{align}\label{masterequationRH}
-\ii\sum_{m=1,2}\dfrac{S_{m}(t)}{2}&\left[L_{m}^{\dagger}L_{m},\rho_{S}(t)\right]+\sum_{m=1,2}\gamma_{m}(t)\left[L_{m}\rho_{S}(t)L_{m}^{\dagger}-\dfrac{1}{2}\left\{L_{m}^{\dagger}L_{m},\rho_{S}(t)\right\}\right]\nonumber\\
=&-\ii\dfrac{S_{1}(t)}{2}\begin{pmatrix}
		0 & r_{1}(t)r_{2}^{*}(t) & r_{1}(t)c_{00}^{*}\\
		-r_{1}^{*}(t)r_{2}(t) & 0 & 0\\
		-r_{1}^{*}(t)c_{00} & 0 & 0
	\end{pmatrix}-\ii\dfrac{S_{2}(t)}{2}\begin{pmatrix}
		0 & -r_{1}(t)r_{2}^{*}(t) & 0\\
		r_{1}^{*}(t)r_{2}(t) & 0 & r_{2}(t)c_{00}^{*}\\
		0 & -r_{2}^{*}(t)c_{00} & 0
	\end{pmatrix}\nonumber\\
 	&-\gamma_1(t)\left[\begin{pmatrix}
	0 & 0 & 0\\
	0 & 0 & 0\\
	0 & 0 & \lvert r_{1}(t)\rvert^2
	\end{pmatrix}-\dfrac{1}{2}\begin{pmatrix}
		2\lvert r_{1}(t)\rvert^2 & r_{1}(t)r_{2}^{*}(t) & r_{1}(t)c_{00}^{*}\\
		r_{1}^{*}(t)r_{2}(t) & 0 & 0\\
		r_{1}^{*}(t)c_{00} & 0 & 0
	\end{pmatrix}\right]\nonumber\\
 &-\gamma_2(t)\left[\begin{pmatrix}
		0 & 0 & 0\\
		0 & 0 & 0\\
		0 & 0 & \lvert r_{2}(t)\rvert^2
	\end{pmatrix}-\dfrac{1}{2}\begin{pmatrix}
		0 & r_{1}(t)r_{2}^{*}(t) & 0\\
		r_{1}^{*}(t)r_{2}(t) & 2\lvert r_{2}(t)\rvert^2 & r_{2}(t)c_{00}^{*}\\
		0 & r_{2}^{*}(t)c_{00} & 0
	\end{pmatrix}\right].
\end{align}
\end{widetext}
Comparing \eqref{masterequationLH} and \eqref{masterequationRH},  it follows that the reduced system dynamics satisfies the master equation
\begin{multline}\label{masterequation}
\dfrac{d\rho_{S}(t)}{dt}=-\ii\sum_{m=1,2}\dfrac{S_{m}(t)}{2}\left[L_{m}^{\dagger}L_{m},\rho_{S}(t)\right]\\
+\sum_{m=1,2}\gamma_{m}(t)\left[L_{m}\rho_{S}(t)L_{m}^{\dagger}-\dfrac{1}{2}\left\{L_{m}^{\dagger}L_{m},\rho_{S}(t)\right\}\right],
\end{multline}
with the Lamb shifts and decay rates
\begin{equation}\label{lambShifts}
    \begin{cases}
        S_{1}(t)=-2\,\text{Im}\left[\dfrac{\dot{r}_{1}(t)}{r_{1}(t)}\right],\\[8pt]
        S_{2}(t)=-2\,\text{Im}\left[\dfrac{\dot{r}_{2}(t)}{r_{2}(t)}\right],
    \end{cases}
\end{equation}
\begin{equation}\label{decayRates}
    \begin{cases}
        \gamma_{1}(t)=-2\,\text{Re}\left[\dfrac{\dot{r}_{1}(t)}{r_{1}(t)}\right],\\[8pt]
        \gamma_{2}(t)=-2\,\text{Re}\left[\dfrac{\dot{r}_{2}(t)}{r_{2}(t)}\right].
    \end{cases}
\end{equation}


\setcounter{secnumdepth}{2}
\setcounter{equation}{0}
\setcounter{figure}{0}
\renewcommand{\theequation}{B\arabic{equation}}
\renewcommand{\thetable}{B\arabic{table}}

\section{Uncorrelated Dynamics}
If we expand the jump operators $L_{1,2}$ in terms of the $\sigma_{-}^{(1,2)}$, we may rewrite the master equation \eqref{masterequation} as
\begin{equation}\label{masterequationcorruncorr}
    \dfrac{d\rho_{S}(t)}{dt}=\mathcal{L}^{\rm uc}_{t}[\rho_{S}(t)]+\mathcal{L}^{\rm c}_{t}[\rho_{S}(t)],
\end{equation}
with
\begin{multline}\label{uncorrliouvillian}
\mathcal{L}^{\rm uc}_{t}[\rho_{S}(t)]=-\ii\dfrac{S_{\rm uc}(t)}{2}\sum_{j=1,2}[\sigma^{(j)}_{+}\sigma^{(j)}_{-},\rho_{S}(t)]\\
+\gamma_{\rm uc}(t)\sum_{j=1,2}\left(\sigma^{(j)}_{-}\rho_{S}(t)\sigma^{(j)}_{+}-\dfrac{1}{2}\left\{\sigma^{(j)}_{+}\sigma^{(j)}_{-},\rho_{S}(t)\right\}\right),
\end{multline}
where $S_{\rm uc}(t)=[S_{1}(t)+S_{2}(t)]/2$ and $\gamma_{\rm uc}(t)=[\gamma_{1}(t)+\gamma_{2}(t)]/2$. It is clear that the remaining part $\mathcal{L}_{t}^{\rm c}$ is responsible for the correlations among the atoms. Therefore, the uncorrelated part of the dynamics is
\begin{equation}\label{masterequationuncorr}
    \dfrac{d\rho^{\rm uc}_{S}(t)}{dt}=\mathcal{L}^{\rm uc}_{t}[\rho^{\rm uc}_{S}(t)],
\end{equation}
We can see from the form of \eqref{uncorrliouvillian} that in this case each atom evolves independently, with the evolution for each one given by 
\begin{equation}\label{rhouncorr1atom}
    \rho^{\rm uc}_{\rm atom}(t)=\begin{pmatrix}
        [\rho^{\rm uc}_{\rm atom}(0)]_{11}|u(t)|^2 & [\rho^{\rm uc}_{\rm atom}(0)]_{10}u(t)\\[10pt]
        [\rho^{\rm uc}_{\rm atom}(0)]_{01}u^{*}(t) & 1-[\rho^{\rm uc}_{\rm atom}(0)]_{11}|u(t)|^2
    \end{pmatrix},
\end{equation}
where $u(t):=\exp\left\{-\frac{1}{2}\int_0^{t}ds[\gamma_{\rm uc}(s)+\ii S_{\rm uc}(s)]\right\}$. From \eqref{masterequationuncorr} we can also calculate
\begin{equation}\label{guncorr}
    g_{\rm uc}(t)=\dfrac{1}{3}\left[|\gamma_{1}(t)+\gamma_{2}(t)|-\gamma_{1}(t)-\gamma_{2}(t)\right].
\end{equation}
Consequently, we are able to compute the measure $\tilde{\mathcal{N}}$ in the case of uncorrelated dynamics.

\setcounter{secnumdepth}{2}
\setcounter{equation}{0}
\setcounter{figure}{0}
\renewcommand{\theequation}{C\arabic{equation}}
\renewcommand{\thetable}{C\arabic{table}}

\section{The limit $\tilde{\mathcal{N}}\rvert_{d\downarrow0}$}
Since $g(t)$ depends continuously on the interatomic distance $d$, we can use its exact expression for $d=0$ in the small distance limit $d\downarrow0$ in the non-Markovianity measure $\tilde{\mathcal{N}}$. The fidelity, however, is discontinuous at $d=0$ when considering $t\rightarrow\infty$, since 
\begin{align}
    \lim_{t\rightarrow\infty}r_{2}(t)=0, \quad \text{for } d>0,\\
    \lim_{t\rightarrow\infty}r_{2}(t)=1, \quad \text{for } d=0.
\end{align}
Specifically, at $d=0$ the fidelity approaches 1 when $r_{1}(t)$ approaches 0, as $r_{2}(t)$ is trivially in its steady state value for any $t$. On the other hand, when considering an arbitrarily small distance $d$, the fidelity will remain at an approximately constant non-vanishing value for all times greater than the time $\tau_1$ that $r_{1}(t)$ takes to relax. After these considerations, we can write
\begin{align}
    \lim_{d\downarrow0}&\lim_{t_{\rm end}\rightarrow\infty}\int_0^{t_{\rm end}}\log F(s)ds\nonumber\\
       &=\lim_{d\downarrow0}\lim_{t_{\rm end}\rightarrow\infty}\int_0^{\tau_1}\log F(s)ds+\int_{\tau_1}^{t_{\rm end}}\log F(s)ds\nonumber\\
       &=\lim_{d\downarrow0}\lim_{t_{\rm end}\rightarrow\infty}\int_0^{\tau_1}\log F(s)ds+\log F(\tau_1) (t_{\rm end}-\tau_1),
\end{align}
and similarly,
\begin{align}
       \lim_{d\downarrow0}\lim_{t_{\rm end}\rightarrow\infty}&\int_0^{t_{\rm end}}\sqrt{g(s)}\log F(s)ds \qquad \quad \nonumber \\
       &=\lim_{d\downarrow0}\lim_{t_{\rm end}\rightarrow\infty}\int_0^{\tau_1}\sqrt{g_0(s)} \log F(s)ds  \qquad  \nonumber\\
       +\int_{\tau_1}^{t_{\rm end}} \sqrt{g_0(s)}\log F(s)ds\span \omit \nonumber\\
    &=\lim_{d\downarrow0}\lim_{t_{\rm end}\rightarrow\infty}\int_0^{\tau_1}\sqrt{g_0(s)}\log F(s)ds\nonumber \\
    +\log F(\tau_1) \int_{\tau_1}^{t_{\rm end}} \sqrt{g_0(s)}ds.\span \omit\nonumber \\[-4pt]
\end{align}
Therefore
\begin{align}
	\tilde{\mathcal{N}}\rvert_{d\downarrow0}&=\lim_{d\downarrow0}\lim_{t_{\rm end}\rightarrow\infty}\left[\dfrac{\int_{0}^{t_{\rm end}}\log F(s)\sqrt{g_{0}(s)}ds}{\int_{0}^{t_{\rm end}}\log F(s)ds}\right]^{2}\nonumber\\
 &=\left[\lim_{t_{\rm end}\rightarrow\infty}\dfrac{\int_{\tau_1}^{t_{\rm end}}\sqrt{g_{0}(s)}ds}{t_{\rm end}-\tau_1}\right]^{2} \nonumber\\
 &=\left[\lim_{t_{\rm end}\rightarrow\infty}\dfrac{\int_{0}^{t_{\rm end}}\sqrt{g_{0}(s)}ds}{t_{\rm end}}\right]^{2},
\end{align}
where we have used the fact that $\sqrt{g_{0}(t)}$ is a periodic integrable function of $t$. Given that in the non-Markovian case
\begin{equation}\label{gd0}
		g_{0}(t)=\frac{8\gamma_0^2}{3}\left[\dfrac{1}{\lambda+|\Omega_{2}|\cot\left(\frac{|\Omega_{2}| t}{2}\right)}\right]^-,
\end{equation}
we can see that $\sqrt{g_0(t)}$ is zero everywhere except inside the intervals $[\tau^{n}_{i},\tau^{n}_{f}]$ with $\tau^{n}_{i}=T[n-\frac{1}{\pi}\cot^{-1}(\frac{\lambda}{|\Omega_2|})]$ and $\tau^{n}_{f}=nT$, where $T=\frac{2\pi}{|\Omega_2|}$ is the period of the function and $n=1,2,3,...$ Thus, we can rewrite the limit as
\begin{multline}
	\tilde{\mathcal{N}}\rvert_{d\downarrow0}=\left[\lim_{n\rightarrow\infty}\dfrac{\int_{0}^{\tau^{n}_{f}}\sqrt{g_{0}(s)}ds}{\tau^{n}_{f}}\right]^{2}\\
=\left[\lim_{n\rightarrow\infty}\dfrac{nI_{0}}{nT}\right]^{2}=\left[\dfrac{I_{0}}{T}\right]^{2},
\end{multline}
where we have defined $I_0=\int_{\tau^{1}_{i}}^{\tau^{1}_{f}}\sqrt{g_{0}(s)}ds$. This can be calculated exactly, yielding
\begin{equation}\label{gd0}
		I_0=\frac{2\pi}{\sqrt{6\gamma_0+3\lambda}}.
\end{equation}
Therefore, inserting this result and the expression for $|\Omega_2|$ in the limit, we find
\begin{equation}\label{gd0}
		\tilde{\mathcal{N}}\rvert_{d\downarrow0}=\frac{1}{3}(2\gamma_0-\lambda).
\end{equation}

\setcounter{secnumdepth}{2}
\setcounter{equation}{0}
\setcounter{figure}{0}
\renewcommand{\theequation}{D\arabic{equation}}
\renewcommand{\thetable}{D\arabic{table}}
\section{Exact dynamics of $N$ atoms at $d=0$}

We consider now the previous model for $N$ identical two-level atoms located at $d=0$. The interaction Hamiltonian is given by
\begin{equation}\label{hamiltonianNat}
    H_{I}(t)=\displaystyle\sum_{\bm{k}}g(\omega_{k})\sum_{n=1}^{N}\left[\sigma_{+}^{(n)}a_{\bm{k}}\text{e}^{-i(\omega_{k}-\omega_0)t}+\text{H.c.}\right].
\end{equation}
This preserves the number of initial excitations. We restrict ourselves to the (0 and) 1 initial excitation subspace, so we have a state at time $t$ of the form
\begin{equation}\label{stateNat}
    \ket{\psi(t)}=c_{0}\ket{0_N}\ket{0}+\sum_{n=1}^{N}c_{n}(t)\ket{1_{n}}\ket{0}+\sum_{\bm{k}}c_{\bm{k}}(t)\ket{0_N}\ket{1_{\bm{k}}},
\end{equation}	
where $\ket{0_N}$ denotes all atoms in the ground state, and \(\ket{1_{n}}\) the \(n\)th atom in the excited state with the others in the ground. As \(\ket{0_N}\ket{0}\) remains invariant, we do not consider it in the following. For the rest, the Schr\"odinger equation gives 
\begin{equation}\label{sysNat}
    \begin{cases}
        \displaystyle \ii\,\dot{c}_{n}(t)=\sum_{\bm{k}}g(\omega_{k})\ee^{-\ii\Delta_{k}t}c_{\bm{k}}(t),\\
        \displaystyle \ii\,\dot{c}_{\bm{k}}(t)=g(\omega_{k})\sum_{n=1}^{N}\ee^{\ii\Delta_{k}t}c_{n}(t).
    \end{cases}
\end{equation}
Since we consider the field initially in the vacuum, following the same procedure as in the case of two atoms, this can be reduced to
\begin{equation}\label{sysNat2}
	\dot{c}_{n}(t)=-\dfrac{\gamma_0^2}{2}\int_{0}^{t}dt'\text{e}^{-\lambda(t-t')}\sum_{m=1}^{N}c_{m}(t'),
\end{equation}
in the continuum limit. In order to solve these equations it is convenient to work in the following basis:
\begin{equation}\label{stateNstdCoeffs}
	\begin{cases}
		\displaystyle\ket{+}=\dfrac{1}{\sqrt{N}}\sum_{n=1}^{N}\ket{1_n},\\
		\displaystyle\ket{-,n}=\dfrac{1}{\sqrt{n(n-1)}}\left[\left(\sum_{m=1}^{n-1}\ket{1_{m}}\right)-(n-1)\ket{1_{n}}\right],
	\end{cases}
\end{equation}
where $n=2,3,...,N$.
In this new basis, state \eqref{stateNat} can be expressed as
\begin{multline}\label{stateNstd}
\ket{\psi(t)}=c_{0}\ket{0_N}\ket{0}+c_{+}(t)\ket{+}\ket{0}\\
+\sum_{n=2}^{N}c_{0,n}(t)\ket{-,n}\ket{0}+\sum_{\bm{k}}c_{\bm{k}}(t)\ket{0_N}\ket{1_{\bm{k}}},
\end{multline}
with
\begin{equation}
	\begin{cases}
		\displaystyle c_{+}(t)=\dfrac{1}{\sqrt{N}}\sum_{n=1}^{N}c_{n}(t),\\
		\displaystyle c_{-,n}(t)=\dfrac{1}{\sqrt{n(n-1)}}\left[\sum_{m=1}^{n-1}c_{m}(t)-(n-1)c_{n}(t)\right].
	\end{cases}
\end{equation}
Taking \(1/\sqrt{N}\sum_{n}\) in \eqref{sysNat2} we obtain
\begin{equation}
	\dot{c}_{+}(t)=-\dfrac{N\gamma_0^2}{2}\int_{0}^{t}dt'\text{e}^{-\lambda(t-t')}c_{+}(t').
\end{equation}
This equation can be easily solved by taking time-derivative or by Laplace transformation, resulting in
\begin{equation}\label{solNat}
	c_{+}(t)=c_{+}(0)\text{e}^{-\frac{\lambda t}{2}}\left[\cosh\left(\dfrac{\Omega_{N}t}{2}\right)+\dfrac{\lambda}{\Omega_{N}}\sinh\left(\dfrac{\Omega_{N}t}{2}\right)\right],
\end{equation}
where $\Omega_{N}=\sqrt{\lambda^2-2N\gamma_0^2}$. For the remaining coefficients, it can be easily seen from \eqref{sysNat2} that $\dot{c}_{-,n}(t)=0$, so the $\ket{-,n}$ states are invariant. Consequently, there is an $N$ dimensional DFS consisting of the states $\{\ket{0_N},\ket{-,n}\}$ with $n=2,3\ldots,N$. 

\setcounter{secnumdepth}{2}
\setcounter{equation}{0}
\setcounter{figure}{0}
\renewcommand{\theequation}{E\arabic{equation}}
\renewcommand{\thetable}{E\arabic{table}}
\section{Exact master equation for $N$ atoms at $d=0$}
The same procedure as in the case of two atoms can now be used to obtain an exact master equation of the system. From \eqref{stateNstd} it follows that the reduced density matrix is given by
\begin{align}
    \rho_{S}(t)&=\Tr_{R}\big[\ket{\psi(t)}\bra{\psi(t)}\big]\nonumber \\
    &=\lvert c_{+}(t)\rvert^2\ket{+}\bra{+}+\sum_{n=2}^{N}c_{+}(t)c^{*}_{-,n}\ket{+}\bra{-,n}\nonumber \\
    &\quad +c_{+}(t)c_{0}^{*}\ket{+}\bra{0_N}+\sum_{n=2}^{N}c_{-,n}c_{+}^{*}(t)\ket{-,n}\bra{+}\nonumber\\
    &\quad +\sum_{m,n=2}^{N}c_{-,n}c^{*}_{-,m}\ket{-,n}\bra{-,m}\nonumber\\
    &\quad+\sum_{n=2}^{N}c_{0}^{*}c_{-,n}\ket{-,n}\bra{0_N}+c_{0}c_{+}^{*}(t)\ket{0_N}\bra{+}\nonumber \\
    &\quad+\sum_{n=2}^{N}c_{0}c^{*}_{-,n}\ket{0_N}\bra{-,n}\nonumber\\
    &\quad +\left(1-\lvert c_{+}(t)\rvert^2-\sum_{n=2}^{N}\lvert c_{0,n}\rvert^2\right)\ket{0_N}\bra{0_N},
\end{align}
where we have used the condition $\Tr\left[\rho_{S}(t)\right]=1$ to remove the dependency on $\bm{k}$. After taking time-derivative we get
\begin{align}\label{meNatLH}
    \dfrac{d\rho_{S}(t)}{dt}&=2\,\left[\dfrac{\dot{c}_{+}(t)}{c_{+}(t)}\right]\Bigg[\lvert c_{+}(t)\rvert^2\left(\ket{+}\bra{+}-\ket{0_N}\bra{0_N}\right)\nonumber \\
    &\quad+\dfrac{1}{2}\sum_{n=2}^{N}c_{+}(t)c^{*}_{-,n}\ket{+}\bra{-,n}+\dfrac{1}{2}c_{+}(t)c_{0}^{*}\ket{+}\bra{0_N}\nonumber \\
    &\quad\left.+\dfrac{1}{2}\sum_{n=2}^{N}c_{-,n}c_{+}^{*}(t)\ket{-,n}\bra{+}+\dfrac{1}{2}c_{0}c_{+}^{*}(t)\ket{0_N}\bra{+}\right],
\end{align}
where we have used the fact that $\dot{c}_{+}(t)/c_{+}(t)\in\mathbb{R}$, as can be seen immediately from \eqref{solNat}. Let us introduce now the collective raising and lowering operators
\begin{equation}
	J_{-}=J_{+}^{\dagger}=\dfrac{1}{\sqrt{N}}\sum_{n=1}^{N}\sigma_{-}^{(n)},
\end{equation}
which satisfy
\begin{align}
	&J_{-}\ket{0_N}=J_{-}\ket{-,n}=0,\\[8pt]
	&J_{-}\ket{+}=\ket{0_N},\\[8pt]
	&J_{+}\ket{0_N}=\ket{+}.
\end{align}
We find that
\begin{align}\label{meNatRH}
    \gamma_{N}&(t)\left(J_{-}\rho_{S}(t)J_{+}-\dfrac{1}{2}\left\{J_{+}J_{-},\rho_{S}(t)\right\}\right)\nonumber\\
    &=\gamma_{N}(t)\Bigg(\lvert c_{+}(t)\rvert^2\ket{0_N}\bra{0_N}-\lvert c_{+}(t)\rvert^2\ket{+}\bra{+}\nonumber\\
    &\quad -\dfrac{1}{2}\sum_{n=2}^{N}c_{+}(t)c^{*}_{-,n}\ket{+}\bra{-,n}-\dfrac{1}{2}c_{+}(t)c_{0}^{*}\ket{+}\bra{0_N}\nonumber\\
    &\quad \left.-\dfrac{1}{2}\sum_{n=2}^{N}c_{-,n}c_{+}^{*}(t)\ket{-,n}\bra{+}-\dfrac{1}{2}c_{0}c_{+}^{*}(t)\ket{0_N}\bra{+}\right).
\end{align}
By comparing \eqref{meNatLH} and \eqref{meNatRH}, it follows that the reduced system dynamics satisfies the master equation
\begin{equation}\label{meNat}
	\dfrac{d\rho_{S}(t)}{dt}=\gamma_{N}(t)\left(J_{-}\rho_{S}(t)J_{+}-\dfrac{1}{2}\left\{J_{+}J_{-},\rho_{S}(t)\right\}\right),
\end{equation}
with  the decay rate
\begin{equation}
	\gamma_{N}(t)=-2\,\left[\dfrac{\dot{c}_{+}(t)}{c_{+}(t)}\right]=\dfrac{2N\gamma_0^2}{\lambda+\Omega_{N}\coth\left(\frac{\Omega_{N}t}{2}\right)}.
\end{equation}

\end{document}